\documentclass[fleqn,twoside]{article}
\usepackage[headings]{mod-espcrc2}
\usepackage{graphicx}

\usepackage{latexsym,amsmath,amssymb,amscd,cite}
\usepackage{hyperref}
\usepackage[all]{xy}

\newcommand{\beq}{\begin{equation}}
\newcommand{\eeq}{\end{equation}}
\newcommand{\bea}{\end{eqnarray}}
\newcommand{\eea}{\end{eqnarray}}
\newcommand{\non}{\nonumber}
\newcommand{\beql}[1]{\beq\label{#1}}

\newcommand{\beqal}[1]{\begin{eqnarray}\label{#1}}
\newcommand{\eeqal}{\end{eqnarray}}

\def\ie{i.e.}

\newcommand{\diag}{{\rm diag}}

\def\del{\partial}

\def\cE{{\mathcal E}}
\def\cJ{{\mathcal J}}
\def\cR{{\mathcal R}}

\def\BC{{\mathbb C}}

\def\BP{{\mathbb P}}

\def\BZ{{\mathbb Z}}
\def\ZZ{{\mathbb Z}}

\def\al{\alpha}
\def\bfone{{\bf 1}}
\def \unity{\bf 1}

\def\weff{{\mathcal W}_{\rm eff}}

\def\BP{{\mathbb P}}
\def\cJ{{\mathcal J}}
\def\cE{{\mathcal E}}
\def\cV{{\mathcal V}}

\title{
Matrix Factorizations, $D$-Branes and their Deformations}

\author{
H. Jockers\address[CERN]{
Theory Division, Physics Dept.\\
 CERN, 1211 Geneva 23, Switzerland
}
and 
W. Lerche\addressmark[CERN]
}
       
\begin{document}

\begin{abstract}
We review in elementary, non-technical terms the description of
topological $B$-type of $D$-branes in terms of boundary Landau-Ginzburg theory, as well as some applications.
\vspace{1pc}
\end{abstract}

\maketitle

\section{INTRODUCTION}

$D$-branes (see e.g. ref.~\cite{Johnson:2000ch}) play an important
r\^ole for understanding certain properties of string and field
theories, as well as for building semi-realistic models. However,
practically all literature on string phenomenology deals with weakly
coupled theories, where compactification radii are large and notions
of classical geometry apply: e.g., supergravity solutions, 
branes wrapping $p$-dimensional cycles, gauge field configurations
on top of branes, etc. All this corresponds
just to the boundary of the parameter space, which (presumably) is
a subset of measure zero of the full string parameter space.

In order to improve the understanding of how string theory behaves
in the main part of its parameter space, we thus need to move away
from the large radius/weak coupling regime. However, naive geometrical
notions, such as a $D$-brane wrapping some $p$-dimensional cycle
of a Calabi-Yau manifold, then start to break down. When distances
become small or curvatures large, quantum corrections tend to blur
notions of classical geometry, such as the dimension of a wrapped
submanifold. Various physical phenomena can arise, like branes can
become unstable and decay in ways that are not visible classically;
orientifold planes can disintegrate; new branches in the moduli
space can open up; new, non-perturbative critical points of the
effective potential can develop; and contrarily, the moduli space
of branes can be obstructed while classically it seems unobstructed
(in other words, a non-perturbative superpotential can be generated).
Recent examples of string phenomenology deep inside the ``bulk"
of the moduli space are given e.g., by refs.~\cite{Becker:2006ks,Diaconescu:2007ah}.

In order to enter the bulk of the moduli space and meaningfully
describe such phenomena, we need to adopt a suitable language for
describing general $D$-brane configurations that goes beyond the
notion of branes wrapping cycles.  For topological $B$-type
\cite{Ooguri:1996ck,HIV,mirbook} $D$-branes, the proper mathematical framework
is a certain enhanced, bounded derived category of coherent sheaves
\cite{Sharpe:1999qz,Douglas:2000gi,Lazaroiu:2001jm,Diaconescu,Aspinwall:2001pu,Distler:2002ym};
via homological mirror symmetry this maps to the category of $A$-type
branes, which wrap Lagrangian cycles of the Fukaya category
\cite{kontsevich,mirbook} or coisotropic $A$-type branes
\cite{Kapustin:2001ij,Kapustin:2003kt,Grange:2004ah}.

This framework treats branes as abstract, not necessarily naive
geometrical objects, but even in the geometrical, large radius limit
it retains more data than the more familiar characterization of
branes in terms of $K$-theory or cohomology (\ie, $RR$ charges).
It thus provides a much sharper description of $D$-branes.  That
is, the category also contains the information about the brane
locations, and other possible gauge bundle moduli.  For instance,
a confi\-guration consisting of an anti-$D0$-brane located at some
point $u_1$ of the compactification manifold, plus a $D0$-brane
located at some other point $u_2$, is trivial from the $K$-theory
point of view, but is a non-trivial object in the categorical
description as long as $u_1\not=u_2$. Obviously, this extra information is crucial for understanding questions such as whether, 
for a given $D$-configuration, deformations are obstructed or not (\ie, what
is the effective superpotential and the moduli space of its flat
directions). Moreover, the language of categories is tailor-made
for addressing questions about stability and bound state formation,
which can be described more physically by tachyon condensation.
Excellent reviews of these matters may be found in
refs.~\cite{Lazaroiu:2003md,Sharpe:2003dr,Aspinwall:2004jr}.

Often physicists associate with derived categories just an abstract
collection of objects (the $D$-branes) and maps (open strings)
between them, and wonder what concrete physical benefit such a
formal picture might provide. Indeed, by merely tracing arrows around a
quiver diagram, all one obtains is a list of possible terms
in the effective superpotential and these terms are usually added up
with unit coefficients. However, there is much more to these maps than just
being pointers between objects: in general they depend on various
parameters like Calabi-Yau and brane moduli, and thus encode valuable extra
information beyond mere combinatorics. Accordingly, superpotential terms
derived from quiver diagrams will in general have pre-factors depending
on the various moduli of the geometry, a fact that is often neglected in
the physics literature.

For $B$-type topological branes, the notions of objects and morphisms
(maps) can, in fact, be easily translated into a very concrete and
useful language more familiar to physicists, namely the language
of two-dimensional topological (twisted $N\!=\!2$ supersymmetric)
boundary Landau-Ginzburg~(LG) models
\cite{Warner:1995ay,Govindarajan:1999js,Govindarajan:2000my}.  As
we will explain in more detail below, the objects, which correspond
to the $D$-branes (respective boundary conditions of open string
world-sheets), are represented by matrix factorizations
\cite{Kapustin:2002bi,Brunner:2003dc}, and the maps between them
are represented by certain matrix valued, moduli dependent open-string
vertex operators.  That this simple physical model faithfully
represents the abstract mathematical notion of the category of
coherent sheaves is highly non-trivial, and has been proven recently
\cite{OrlovB} to quite some generality (see also
\cite{Aspinwall:2006ib,HeHo}).  The key point is that the relevant
category of topological $B$-type $D$-branes on a Calabi-Yau manifold
described by $W=0$, is isomorphic to a certain category of matrix
factorizations of $W$ \cite{KontsU,OrlovA}.  These factorizations
of the from $W\bfone=\cJ\cdot \cE$ then have a direct interpretation
in terms of the boundary Landau-Ginzburg model, where $W$ figures
as the bulk superpotential and $J$ as a boundary superpotential.
Given a bulk Calabi-Yau geometry defined by $W=0$, the specific
choice of $\cJ$ (and consequently, $\cE$) encodes the specific
$D$ brane geometry that the LG model describes.

A great advantage of the LG formulation, over approaches based on
rational boundary CFT \cite{Recknagel:1997sb,Brunner:1999jq}, is
that one can easily study the dependence of physical observables
on moduli, and as well as on relevant deformations. In particular, it allows to
quantitatively study phenomena like tachyon condensation
\cite{Govindarajan:2005im}, and to explicitly compute instanton
corrected, effective superpotentials on world-volumes of intersecting
branes
\cite{Ashok:2004xq,Hori:2004ja,Brunner:2004mt,Govindarajan:2006uy,Knapp:2007kq,WL}.  
Moreover, it allows to address global properties of branes in the
K\"ahler moduli space \cite{Jockers:2006sm}, and more generally
an extension to gauged linear $\sigma$-models allows to interpolate
the brane data among different points in the K\"ahler moduli space \cite{HeHo}.
Some works analyze the connection of matrix factorizations to
rational boundary CFTs at the Gepner point 
\cite{Brunner:2005fv,Brunner:2005pq,Enger:2005jk,Brunner:2006tc}, 
show the relation
of matrix factorizations to homological knot invariants \cite{Gukov:2005qp}, 
or extend matrix factorizations to orientifold models \cite{Hori:2006ic,Brunner:2006yi}.
For works covering other aspects of matrix factorizations see, for example, refs.~\cite{Kapustin:2003rc,Kapustin:2003ga,Herbst:2004jp,Ashok:2004zb,Walcher:2004tx,Dell'Aquila:2005jg,Knapp:2006rd,Knapp:2007kq,Aspinwall:2007cs,Baumgartl:2007an,Brunner:2007qu}.

In these lectures, we will review some of these aspects in simple
terms, leaving more complicated mathematics and technical details
to the original papers. In the next section, we will recall some
basic features of boundary LG theory in relation to matrix
factorizations. In section 3 we will then study deformations of
matrix factorizations and show how obstructions of them lead
very directly to effective superpotentials.  In section 4, we will
turn to discussing in more detail, as a case study, branes on the
elliptic curve. This geometry is well understood both from the
mathematics as well as from the physics side, and it is non-trivial
enough in order to get some idea about how the general case works.
Specifically, we will show how the bundle data $(r(\cE),c_1(\cE),u)$,
that characterize a given brane, are explicitly encoded by certain
properties of the matrices. Moreover we will discuss a simple example
of tachyon condensation.

\section{LG MODELS AND MATRIX FACTORIZATIONS}

We start by briefly reviewing the bulk Landau-Ginzburg model;
for more details about this well-known theory, see for example \cite{mirbook}.
It is easiest written in terms of $d=2$, $N=(2,2)$ superspace language
as follows:
\beql{bulkLG}
S_{LG}=\int\!\! d^2zd\theta^4K(\Phi_i,\bar \Phi_i) + 
\int\!\! d^2zd\theta^2W(\Phi_i)+{\rm c.c.}\ ,
\eeq
Here, $\Phi_i$, $i=1,...,n$ are chiral superfields, which form
reduced supermultiplets that satisfy $\bar D_\pm\Phi_i=0$ 
(in the following we will use the notation $x_i$ when we
consider the fields as complex variables rather than superfields).
Moreover, $K(x_i,\bar x_i)$ is the K\"ahler potential which is a
non-holomorphic function of the LG fields, and by standard
renormalization group arguments it does not play a r\^ole in the
infrared. The superpotential $W(x)$, on the other hand, is a
holomorphic function and thus protected by supersymmetry. For 
a quasi-homogenous superpotential, which means that it uniformly
scales like $W(s^{q_i}x_i)=s W(x_i)$, the theory flows in the
infrared to a superconformal fixed point theory that only depends
on the singularity type of $W$.  Given the $R$-charges $q_i$ of the
LG fields, the central charge of that  SCFT is simply given by
$c=3\sum(1-2 q_i)$. Note that this does certainly not specify the
CFT uniquely, rather $W$ may in general contain continuous parameters
on which the singularity type, and thus the CFT, depends.

The significance of $W$ is that it describes the internal compact
background geometry on which the closed strings propagate. The
clearest and most important geometrical interpretation is when
$n=d+2$ and $\sum q_i=1$: then the hypersurface defined by $W=0$
describes a Calabi-Yau $d$-fold $X_W$ with vanishing first Chern
class.  For example, in Section 4 we will focus on
\beq
W(x_i,a)\ =\ {x_1}^3+{x_2}^3+{x_3}^3-3\, a\,x_1x_2x_3\ .
\eeq
The equation $W(x_i,a)=0$ describes the cubic elliptic curve as a sub-manifold
of $\BC\BP^2$, and $a$ parametrizes its complex structure; more details
about this later.

Such a LG theory can be ``twisted'' by adding a background charge
\cite{mirbook}, to the effect that two of the $N=2$ supercharges
turn into spin zero BRST operators. Here we will consider the
``$B$-type'' twist for which the BRST operators are given by $ \bar
Q_+$ and $\bar Q_-$.  Upon such a twisting, the theory becomes a
topological CFT with a finite dimensional Hilbert space defined by
the non-trivial cohomology of $\bar Q_\pm$
 (which implies that the anti-chiral sector is dropped).
More precisely, the spectrum of physical operators is given by the
set of primary chiral fields, which can be represented by simple
polynomials of the LG field $x$, subject to a truncation condition
given by the gradient of the superpotential. In other words, the
physical spectrum consists of the following polynomial ring:
\beq
\cR\cong\BC[x_i]/[\del_iW(x)=0]\ .
\eeq

We now consider the open string version, namely the LG model
(\ref{bulkLG}) with superpotential $W$ on a Riemann surface with
boundary, \ie, on the disk $D$. The boundary breaks the $N=(2,2)$
supersymmetry of the bulk theory to $N=2$ supersymmetry, and we
will choose ``$B$-type'' boundary conditions \cite{Ooguri:1996ck}
that are compatible with the topological twist.  The surviving
supercharge (rather: BRST operator) can then be taken as $Q_{B}\equiv\bar
Q_++\bar Q_-$.

However, as is typical for supersymmetric Lagrangians, (\ref{bulkLG})
is supersymmetry invariant only up to a total derivative. Therefore,
in the presence of a boundary, supersymmetry is broken by a
boundary term, which in the present context is called ``Warner
term'' \cite{Warner:1995ay}. It has the following form:
\beqal{warnerterm}
Q_{B}\cdot\int_D&&\!\!\!\!\!\!\!\!\!\!\!\!\!\!\!
d^2zd^2\theta\,W(\Phi)\non\\
=&&\!\!\!\!\!\!
\int_Dd^2zd\theta^+d\theta^-\,(\theta^+\del_++\theta^-\del_-)
\,W(\Phi)\non\\
=&&\!\!\!\!\!\!
\int_{\del D}d\sigma d\theta\,W(\Phi\vert_{\del D})\ .
\eeqal
To restore supersymmetry, one may introduce extra degrees of freedom
living on the boundary, and cook up a suitable boundary potential
whose supersymmetry variation cancels (\ref{warnerterm}).  As we
will see, there is in general a large ambiguity in doing so, and
the set of all consistent choices corresponds, essentially,
to the set of all supersymmetric boundary conditions, or topological
$B$-type $D$-branes, that are compatible with a given closed
string geometry $X_W$.

The simplest possibility is to introduce boundary fermions \cite{Warner:1995ay,GovBFermions}
$\pi_a,\,\bar\pi_a$, $a=1,...,k$ for some $k$, which satisfy a
Clifford algebra: $\{\pi_a,\bar\pi_b\}=\delta_{ab}$. These fields fit
into fermionic supermultiplets of the $d=1$, $N=2$ superalgebra,
which can be compactly written as superfields of the form
$\Pi_a\equiv(\pi_a+\theta\ell_a)$, where $\ell_a$ are auxiliary
fields. A peculiarity of such superfields is that they are not
truly chiral but satisfy \cite{Govindarajan:1999js,Govindarajan:2000my}
\beq
\bar D\,\Pi_a\ =\ E_a(\Phi\vert_{\del D})\ ,
\eeq
where $E_a(x)$ are a priori arbitrary polynomials in $x$. 

Given the fermionic boundary superfields, we can then add
the following boundary superpotential term:
\beq
S_\del\ =\ \int_{\del D}d\sigma d\theta\,\sum_a \Pi_aJ_a(\Phi\vert_{\del D})\ ,
\eeq
and one can easily check that its $Q_B$-variation indeed cancels the Warner term
(\ref{warnerterm}) iff  \cite{Kapustin:2002bi,Brunner:2003dc}
\beql{fact1}
\sum_a J_a(x)E_a(x)\ =\ W(x)\ .
\eeq
This simple equation turns out to have important consequences!
Before discussing those, let us first rewrite it by making use of
the fact that the BRST operator associated with the boundary degrees
of freedom can be represented as
\beql{Qbound}
Q(x) = \pi_aJ_a(x)+\bar \pi_aE_a(x) = 
\begin{pmatrix}
0&\cJ(x)\\
\cE(x)&0
\end{pmatrix}_{2^k\times2^k}\!\!\!\!.
\eeq
Here we have represented the $k$ pairs of boundary fermions by generalized,
$2^k\times2^k$ dimensional Pauli matrices. The condition (\ref{fact1}) then turns into
the following matrix equation:
\beql{fact2}
Q^2(x)\ =\ \cJ(x)\cdot\cE(x)\ =\ \cE(x)\cdot\cJ(x)\ =\ W(x)\bfone\ .
\eeq
This equation is, in fact, more fundamental than (\ref{fact1}),
since introducing boundary fermions is not really necessary; using
appropriate superholonomy factors in the path integral
\cite{Herbst:2004ax} one can avoid fermions alltogether
and write everything directly in terms of matrices. This lifts the
restriction on the dimension of the matrices to coincide with the
dimension of a Clifford algebra.

Eq.~(\ref{fact2}) means, essentially, that all the supersymmetric
$B$-type boundary conditions (and thus, $B$-type $D$-branes)
compatible with a closed string background $X_W$ described by $W=0$,
are one-to-one to all the possible factorizations of the superpotential
$W(x)$ into matrices $\cJ$ and $\cE$ that have polynomial entries
in the $x_i$ and otherwise are arbitrary. This is a very
non-trivial mathematical statement, since we know that the $B$-type
$D$-branes are described by a certain category $D^b(Coh(X_W))$ of
coherent sheaves associated with $X_W$, which has in general a very complicated 
structure (in particular for Calabi-Yau threefolds). 
As mentioned in the introduction,
this statement has been proven in \cite{OrlovB,Aspinwall:2006ib,HeHo}
by constructing a category of matrix factorizations associated with
$W$, and showing that it is equivalent to $D^b(Coh(X_W))$.

A key ingredient is that the objects $P$ in this category of matrix
factorizations have a composite structure, essentially determined
by the block off-diagonal pieces $\cJ$ and $\cE$ of $Q$. These can
be viewed as maps between ``constituents'' $p_+$ and $p_-$, so that
one can represent the object $P$ as a complex of the form:\footnote{One
of the main points of the proof is to ``unfold'' this
$\ZZ_2$-graded complex into a $\ZZ$-graded complex that
describes a coherent sheaf in the geometrical, large radius limit.
For details see \cite{OrlovB,Aspinwall:2006ib,HeHo}.}
\beqal{Pdef}
P \cong\ \Bigl(\xymatrix{p_+\ \ar@<0.6ex>[r]^{\cE} &\ p_- \ar@<0.6ex>[l]^{\cJ}}
\Bigl)\ .
\eeqal
This abstract mathematical construction (which is due to Kontsevich
\cite{KontsU}) has a nice physical interpretation \cite{Kapustin:2002bi}
which has been made rigorous in \cite{Herbst:2004ax}: The ``components''
$p_\pm$ correspond to $D$-branes and anti-$D$-branes, and the
presence of the potential $W$ forces them to condense, in the IR
limit, into the $D$-brane configuration that we are interested in to
describe. The maps $\cJ$ and $\cE$, which figure as boundary
potentials and whose presence is required for restoring supersymmetry
(c.f., (\ref{fact2})), are viewed as a tachyon field configuration
that effects this condensation.

This simple picture also allows us to easily understand what the geometry of the 
resulting $D$-brane is in terms of $\cJ$ and $\cE$. As a toy example, consider 
the situation with just one LG field. Then one can write
\beql{simplestJ}
\cJ(x)=J(x)=\prod_{a=1}^L(x-u_a)\ ,
\eeq
which describes a system of $L$ $D0$ branes (located at the zeros
$u_a$) that arises via condensation of a $D2$-$\bar{D2}$
system. This system does not completely annihilate because the
tachyon configuration is topologically non-trivial (this is an idea
originally due to Sen \cite{Sen:1998sm}). The first Chern class of
the tachyon condensate is given by the winding number of the map
$J(x)$, and thus is equal to $L$. Via the usual $K$-theory arguments,
$c_1=L$ indeed measures the number of $D0$-branes.

\begin{figure}[htb]
\vspace{2pt}
\includegraphics[width=15pc]{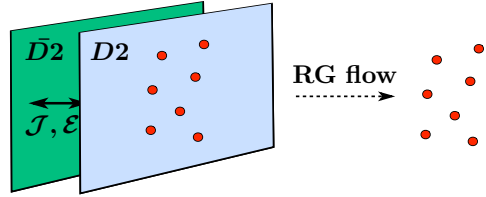}
\caption{Tachyon condensation of a $D2$-$\bar{D2}$ system, triggered by
the boundary potentials $\cJ$ and $\cE$, leads in the infrared to
a residual $D0$-brane configuration described by the matrix factorization
$\cJ\cdot\cE=W$.}
\label{fig:RGflow}
\end{figure}

\goodbreak The construction is analogous for general $B$-type $D$-brane
geometries.  At large radius where semi-classical geometry applies,
they can be thought of as vector bundles (or sheaf generalizations
thereof), localized on holomorphic sub-manifolds of $X_W$.
Again, the maps $\cJ$ and $\cE$ encode certain topological data
such as Chern classes of bundles and sheaves on the space $X_W$,
and these Chern classes directly translate to the $RR$-charges of
the $D$-brane configuration that remains after tachyon condensation.
We will exemplify this for the elliptic curve in
Section~4.

Note, however, that this construction is more general than just
$K$-theory, \ie, $RR$-charges. The maps $\cJ$ and $\cE$ may depend also
on continuous geometrical data, which correspond to deformations of the
$D$-brane configuration; in the previous example these are the $D0$
brane locations~$u_a$ in (\ref{simplestJ}). In general, two configurations which
have different values of the moduli but otherwise are the same,
correspond to different objects in the category, while they are
indistinguishable from the $K$-theory viewpoint.  For certain values,
the maps may degenerate and lead to interesting physical effects,
like jumps in the spectrum of physical operators. Moreover, the
moduli dependence of the boundary potentials $\cJ$ and $\cE$ trickles
down to the correlation functions of the LG theory, and eventually
leads to a (sometimes explicitly computable) moduli dependence of
the effective action 
\cite{Ashok:2004xq,Hori:2004ja,Brunner:2004mt,Govindarajan:2006uy,Knapp:2007kq,WL}.

In order to address this kind of questions, we first of all
need to define the spectrum of the physical open-string states.
Again, these are viewed as maps, but now as maps between the composite objects
(\ref{Pdef}). For open strings beginning and ending on the same
brane $P_a$, which correspond to ``boundary preserving'' vertex
operators $\Psi_{a}\!=\!\Psi_{aa}(u_a)$, the spectrum is simply
given by the non-trivial BRST cohomology of $Q_a\!=\!Q_{a}(u_a)$.
Mathematically speaking, they are associated with maps from the
given brane onto itself, and the cohomology problem is best
characterized by writing $\Psi_{a}\in{\rm Ext}^s(P_a,P_a)$, because
$\Psi_a$ can be viewed as an extension group element at some grade $s$
(for details see e.g.,~\cite{Katz,Aspinwall:2004jr}).

On the other hand, open strings that go between
different branes are described by ``boundary changing'' operators:
$\Psi_{ab}\!=\!\Psi_{ab}(u_a,u_b)$; these map between different
matrix factorizations and are thus in general given by non-square
matrices. For those, the proper notion of being BRST closed is:
\beql{Qaction}
\Big[Q,\Psi_{ab}\Big]\ \equiv\ Q_a\cdot
\Psi_{ab}-(-)^s \Psi_{ab}\cdot Q_b\ =\ 0\ ,
\eeq
and analogously for being BRST exact 
(the sign reflects the statistics of $\Psi_{ab}$).
Again, the cohomology problem has a well-recognized mathematical
meaning, namely in terms of extension groups
$\Psi_{ab}\in {\rm Ext}^s(P_a,P_b)$ between the two $D$-branes $P_a$
and $P_b$. Such operators are localized at
the intersections of the $D$-branes, and
accordingly their (net) number is given by the intersection number
between the branes which can be written as a topological index:
$\chi_{ab}\!=\!{\rm Tr}_{ab}(-1)^p {\rm dim\,Ext}^p(P_a,P_b)$.

\begin{figure}[htb]
\includegraphics[width=18pc]{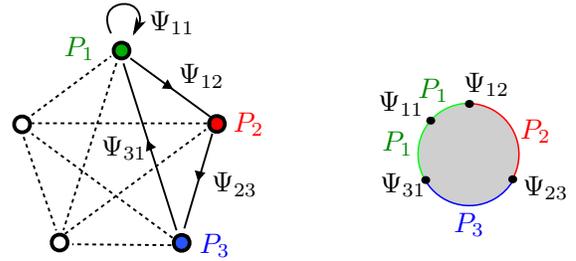}%
\vspace{-15pt}%
\caption{Left: Quiver diagram, where the nodes correspond 
to $D$-branes and the arrows to open strings mapping between them. 
Right: Disk world sheet with corresponding boundary preserving and 
boundary changing vertex operator insertions.}
\label{fig:quiver}
\end{figure}

One often represents the open string spectrum of a brane configuration
in terms of a ``quiver diagram''. See for example 
Fig.~\ref{fig:quiver} which exhibits the r\^ole of boundary changing
and boundary preserving maps, both from the space-time as well as the
world-sheet perspective. 

The simplest of all factorizations is the trivial factorization,
where $J_a=1$ and $E_a=W$. One can easily see that there is no
non-trivial open-string cohomology associated with it (not even the open-string
vacuum exists), and this means that there is no $D$-brane present at
all. This corresponds to the total annihilation of the $D\bar D$
pair due to a topologically trivial tachyon configuration. Brane
configurations differing by such trivial brane-antibrane pairs are
physically equivalent, and correspondingly in the mathematical
formulation, the category is modded out by such "perfect complexes".
For LG models this means that two factorizations of different
dimensions that are related via the addition or removal of matrix blocks of
the form $(\bfone,W\bfone)$, are equivalent.

What happens if we swap $\cJ$ and $\cE$? While
obviously the factorization of $W$ stays invariant, 
$\cJ$ and $\cE$ enter differently in the LG lagrangian.
It turns out that $J_P\to-E_P$ and  $E_P\to-J_P$ maps a
brane $P$ into its anti-brane $\bar P$ (in the math literature the
anti-object is commonly denoted by $P[1]$). A way to see this is
to show that such a $P\bar P$ configuration can completely annihilate
under tachyon condensation.

We will outline this in the next section, but before that we 
introduce another feature that we are going to need, namely that
of gauge symmetries. Indeed a factorization (\ref{fact2}) is invariant
under local gauge transformations of the form:
\beql{gauge}
\begin{aligned}
\cJ(x) \ & \rightarrow  \ U_L(x)\,\cJ(x)\,U_R(x) \ ,\\
\cE(x) \ & \rightarrow  \ U_R^{-1}(x)\,\cE(x)\,U_L^{-1}(x)\ ,
\end{aligned}
\eeq
for polynomial matrices $U_{L,R}(x)$ that are invertible over
$\BC[x_i]$.  In particular, this means we can do arbitrary
row and column reduction operations on $\cJ$ (respectively, $\cE$), as
long as we do the corresponding inverse operations on $\cE$
(respectively, $\cJ$). This is an important tool for exhibiting the
nature of a given factorization, for example a trivial block structure
of the form $(\bfone,W\bfone)$ may not be manifestly visible in a
random gauge. It is therefore useful for determining the outcome
of a condensation process.

\section{DEFORMATIONS}

\subsection{Bulk and boundary moduli}

One of the main advantages of the formulation
of $B$-type of $D$-branes via matrix LG models over other approaches,
is that it is easy to incorporate continuous deformations.  It allows to study
their effects on the spectrum and as well to derive the effective
potential induced on them.  As we will show, one can easily implement ideas of
abstract deformation theory; this is useful because when
deformations are obstructed and an effective potential is generated,
the theory is off-shell away from the critical locus of the effective
potential, and therefore we need to have a formalism that is robust
enough to deal with this.

We will be interested in the following three classes of perturbations
of the open string $B$-model: First, there are the usual complex
structure perturbations of the bulk geometry:
\beql{bulkpert}
W(x)\ \longrightarrow\ W(x,t)\ =\ W_0(x)+ t_i\phi_i+...
\eeq
where the dots indicate that generically we will consider also higher orders in the perturbation. 

Then there are analogous perturbations in the open string, boundary sector.  Here
we distinguish, as before, boundary preserving and boundary changing
perturbations. The latter arise for multiple, intersecting
brane configuration, and we will discuss such perturbations further below.
The boundary preserving deformations 
\beql{bppert}
Q(x)\ \longrightarrow\ Q(x,u)\ =\ Q_0(x)+ u_a\Psi_a+...
\eeq
are tied to a single brane and 
typically correspond to location moduli.\footnote{Note that
because the actual perturbation one adds to the action is of the
form $u_a\int dx\{G^-,\Psi_a\}$, bosonic perturbation parameters
are related to fermionic vertex operators and vice versa. 
Only the bosonic parameters are natural physical deformations
of the theory and will be considered below.} More precisely,
from the supersymmetric GSO projection we know that the bulk operators
and boundary preserving boundary operators have integral charges;
specifically with ``moduli'' we refer to operators with $R$-charge
$q=2$ in the bulk, or $q=1$ on the boundary, so that $t_i$ and
$u_\al$ are dimensionless and can appear in correlation functions
in a non-polynomial way. Strictly speaking, because an effective
potential will be generically generated, some or all of these
parameters won't be true moduli but will be constrained.

\begin{figure}[htb]
\centerline{\includegraphics[width=12pc]{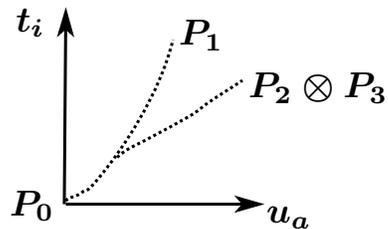}}%
\vspace{-15pt}%
\caption{The supersymmetry preserving deformations form a sub-locus
of the joint open/closed string deformations space. It coincides with the
locus of matrix factorization which is also the critical locus of
the effective superpotential. Along the sub-locus, the brane content and the
open string cohomology may or may not vary, depending on whether the
boundary perturbation is relevant or marginal.
}
\label{fig:loci}
\end{figure}

Indeed, it is obvious that generic perturbations 
(\ref{bulkpert},\ref{bppert}) will
spoil a given factorization and thus lead away from a
supersymmetric configuration.  The true moduli comprise precisely
those deformations that preserve the factorization, \ie, $Q(u)^2=W(t)$,
and these form a sub-locus of the deformation space, as is schematically
shown in Fig.~\ref{fig:loci}.

As is familiar from other instances (for example from
\cite{Cachazo:2002pr}), factorized geometries tend to be related
to critical loci of effective superpotentials, and as we will see,
this is also the case here. This has been discussed and explicitly
worked out in detail \cite{Herbst:2004zm,Knapp:2006rd} for the
topological minimal models as special examples, and more general
discussions in the context of matrix factorizations have been given
in refs.~\cite{Ashok:2004xq,Hori:2004zd,Diaconescu:2006id}. All
this is just a specific realization of some general deformation
theoretical ideas, readable accounts of which are given, for example,
by \cite{Lazaroiu:2001nm,KajiuraAX,FukayaAA}.  We will give here
only a crude presentation of some of the underlying ideology, and
refer the reader to those papers for precise definitions and more
details.

The basic principle is very simple: start with a perturbation of
the BRST operator by a cohomology element as in (\ref{bppert}),
\ie, $Q=Q_0+Q_1$, with $Q_1=u_a\Psi_a\in {\rm Ext}^1(P,P)$. This will in
general spoil factorization at the second order level, \ie,
schematically:
\beql{secondorder}
Q^2-W\ =\ u^2 \{\,\Psi,\Psi\,\}\  \equiv:\ \lambda_2(\Psi,\Psi)\ .
\eeq
One can try to cancel this by introducing a second order correction, 
\ie, by adding
\beq
Q_2\ =\ -U\lambda_2(\Psi,\Psi)\ ,
\eeq
where $U$ is a charge $-1$ operator that is the conjugate or
``inverse'' of $Q$ in the sense that $U^2=0$ and $\Pi=1-\{Q,U\}$
is a projector on physical states. There is an ambiguity in this
definition, and we fix it by requiring a Siegel type of gauge by
requiring $U\Psi=0$ on all physical states.  Thus, roughly speaking,
as long $\lambda_2(\Psi,\Psi)$ is BRST exact, we can use $Q_2$ to
cancel the RHS of (\ref{secondorder}); however then a third order
term, $m_3(\Psi^{\otimes3})$, is induced. But the procedure may be
iterated and eventually yields a solution for $Q=Q_0+\sum_{n>0}^\infty Q_n$
given by:
\beq
Q_n\ =\ -U\cdot\lambda_n(\Psi^{\otimes n})\ .
\eeq
Here $\lambda_n$ are certain higher ``Massey'' products which are recursively defined 
as follows (see e.g., refs.~\cite{FukayaAA,Kajiura:2005sn} for reviews):
\beqal{lambdadefs}
&&\lambda_n(\Psi_1,...,\Psi_n)
=\\
&&(-)^{n-1}[U\lambda_{n-1}(\Psi_1,...,\Psi_{n-1})]\cdot\Psi_n
\non\\
&&- 
(-)^{n|\Psi_1|}\Psi_1\cdot U\lambda_{n-1}(\Psi_2...,\Psi_{n})
\non\\
&&-\sum_{{l+k=n\atop k,l\geq2}}(-)^S
[U\lambda_k(\Psi_1...,\Psi_{k})]\cdot
[U\lambda_l(\Psi_{k+1}...,\Psi_{n})]\non
\eeqal
where $|\Psi|\in\{0,1\}$ denotes the statistics degree of a field,
and $(-)^S$ is some complicated sign.  They form part of an $A_\infty$
structure that is inherent to any open string theory on the disk,
and their recursive structure can be interpreted in terms of tree
level Feynman diagrams associated with open string field theory
\cite{Lazaroiu:2001nm,KajiuraAX,FukayaAA}; this just reflects the
inherent off-shell nature of this formalism which goes beyond
conformal field theory.

However, the procedure goes through only if $Q$ can be inverted at
every step -- otherwise it truncates, signaling an obstruction
against the given perturbation at that order. Obviously $Q$ cannot
be inverted if some $\lambda_n$ is not BRST exact, \ie, if it is a
non-trivial cohomology element.\footnote{Noting that the degree $s$
of $\Psi\in{\rm Ext}^1(P,P)$ equals to one, it follows that the
degree of all the $\lambda_n$ is two, which means that abstractly
speaking, the obstructions are measured by classes in ${\rm
Ext}^2(P,P)$). The pattern continues: ${\rm Ext}^3(P,P)$ measures
obstructions to obstructions, and so on 
\cite{Katz,Berenstein:2002fi,Aspinwall:2004jr}.} The condition for a
perturbation to be unobstructed, and thus to preserve the matrix
factorization of $W$, is
\beql{mcond}
\sum m_n(\Psi^{\otimes n})\ \equiv\ \sum\Pi\,\lambda_n (\Psi^{\otimes n})\ =\ 0\ .
\eeq
Taking the moduli dependence of the $\lambda_n$ into account,
this singles out a sub-locus of the full parameter space.

In deformation theory, the equations (\ref{mcond}) are known as generalized
Maurer-Cartan equations \cite{Lazaroiu:2001nm,KajiuraAX,FukayaAA}. 
They can also be interpreted as equations
of motion of an open string field theory of Chern-Simons type,
and within this context they can be interpreted as the critical
locus of an effective superpotential. That is, from its definition
as generating function of disk correlators, the latter is given by
\beql{Weff}
\weff(t,u)\ =\ \sum_{n=2}^\infty \,\frac1{n+1}\,\langle\Psi,m_n(\Psi^{\otimes n})\rangle\ ,
\eeq
where $\langle\ ,\  \rangle$ is a suitable inner product on the
off-shell Hilbert space, and certain cyclicity properties of the
$m_n$ are implicitly assumed to hold. Eqs.~(\ref{mcond}) are then
simply the equations of motion associated with $\weff(t,u)$.

There is one other important aspect to note, concerning the uniqueness
of the effective potential. We already noted that the inversion of
$Q$ is ambiguous, because one can always add BRST closed pieces to
a BRST non-invariant one. One may fix this ambiguity by adopting
the Siegel type gauge $U(\Psi)=0$ as above, but different gauges
are equally well possible and lead to different results for
$\weff(t,u)$.  However, one can show \cite{Lazaroiu:2001nm} that
the physically relevant data, \ie, the critical locus of $\weff(t,u)$,
remain (essentially) invariant, while the directions in field space
off the critical locus are ambiguous; this reflects that a given
physical theory can have different off-shell completions. More
precisely speaking, all those data are determined only up to field
redefinitions, and that's almost the best one can hope for a
topological field theory. Namely, in order to pin down the
parametrization one would need to know the kinetic terms of the
effective action as well, but these are not available in a TFT
corresponding to $N=1$ space-time supersymmetry.

These statements apply to generic boundary perturbations, which are
typically obstructed.  However, in certain few circumstances, the
open moduli space is flat and this can then be used to treat it
similar to the closed string moduli space, \ie, define preferred
flat coordinates on it.  An example are $D0$ branes on the elliptic
curve which we will discuss below; the moduli space of a $D0$ brane
on some manifold $X_W$ is given by $X_W$ itself, which is flat for
the torus. In contrast, $D0$ branes on a Calabi-Yau threefold
certainly don't have a flat moduli space.

\subsection{Relevant boundary deformations: tachyon condensation}

As said before, there is another class of perturbations in the open
string sector, namely by boundary changing operators, $\Psi_{ab}$,
acting between branes $P_a$ and $P_b$. They have quite different
properties as compared to the boundary preserving moduli.  In
particular, their charges depend on the difference of the `grades'
of the branes they couple to \cite{Douglas:2000gi}. These grades
change continuously when we vary the K\"ahler moduli
\cite{Douglas:2000qw,Aspinwall:2001dz}, and therefore the charges of the open strings
change too. This effect is of course not visible
in the topological $B$-model, which does not depend on K\"ahler
moduli.  In the un-twisted physical theory, however, the K\"ahler moduli do matter and the masses of the open strings will depend on them. Depending on the K\"ahler parameters,
the open strings may become tachyonic, and this is why we will denote
the associated deformation parameters by $T=T_{ab}$.

Given two matrix factorizations representing branes $P_a$ and
$P_b$, and a cohomologically non-trivial fermionic boundary changing
operator $\Psi_{ab}$, we can form a composite matrix factorization
of the form
\begin{equation} \label{eq:MFcone}
    Q_{\rm C} \, = \, 
       \begin{pmatrix} 
            Q_a & T\,\Psi_{ab} \\ 
            0 & Q_b 
       \end{pmatrix} \ .
\end{equation}
It is easy to check with eq.~\eqref{Qaction} that the new matrix
factorization $Q_{\rm C}$ fulfills for any value of the coupling
$T$ the relation $Q_{\rm C}^2=W\bfone$. However, compared to
deformations associated to open-string moduli there is an important
difference. In the topological theory the coupling $T$ does not
change the composite matrix factorization, $Q_{\rm C}$, in a
continuous manner. Instead there are only two distinct choices,
namely either the coupling $T$ is zero, {\it i.e.} no composite is
formed, or the coupling $T$ is non-zero and a composite is formed.
Different non-zero values of the coupling $T$, however, lead to
gauge-equivalent composites, which all represent the very same brane
configuration. Hence, in the topological theory the coupling $T$
gives only rise to a discrete choice, whereas an open-string modulus $u$
parametrizes a continuous family of inequivalent matrix factorizations.

This reminds us that we should be careful about the difference
between topological tachyon condensation and tachyon condensation
in the physical theory. We cannot decide within the topological
sector whether a composite is stable in the underlying physical
theory: this depends on whether the charge of $\Psi$ is less than
one or not, \ie, whether $\Psi$ is a relevant operator or not 
in the physical theory.
Stability of a composite is indeed a complicated concept due its
dependence on the K\"ahler moduli \cite{Douglas:2000qw,Aspinwall:2001dz}.
Namely, in some region of the
K\"ahler moduli space the formation of the composite is energetically
favorable and the coupling $T$ acquires a non-zero (K\"ahler
moduli dependent) vacuum expectation value there.  Then a bound
state represented by the matrix factorization $Q_{\rm C}$ is
formed through tachyon condensation in the physical theory.  However,
in other regions of the K\"ahler moduli space, $T=0$ may be the
energetically favored vacuum expectation value, and the composite
is unstable against the decay  into its constituents, $P_a$ and $P_b$.
For extensive treatments of the notion of $D$-brane stability we
recommend to the reader the
refs.~\cite{Douglas:2000gi,Douglas:2000qw,Denef:2000nb,Aspinwall:2001pu,Aspinwall:2001dz}.

As simplest possible example we consider, as advertized, a generic brane
system given by a matrix factorization (\ref{eq:MFcone}) for which
\beq
Q_a=\begin{pmatrix}
 0&\cJ\\
\cE&0
\end{pmatrix},\ 
Q_b=-\begin{pmatrix}
 0&\cE\\
\cJ&0
\end{pmatrix},\ 
\Psi_{ab}=\begin{pmatrix}
 0&{\bf 1}\\
{\bf 1}&0
\end{pmatrix}.
\eeq
It is easy to check that $\Psi_{ab}$ is a non-trivial cohomology
element, satisfying (\ref{Qaction}) while being non-exact. It is
also easy to see that for $T\not=0$ the factorization can
be brought via gauge transformations (\ref{gauge}) to the form
\beq
    Q_{\rm C} \, = \, 
\begin{pmatrix} 
0&0&0&1\\
0&0&-W&0\\
0&-1&0&0&\\
W&0&0&0&
 \end{pmatrix}\ ,
\eeq
which, according to what we said before, represents the trivial
brane configuration. This justifies a posteriori the claim that
$Q_b$ represents the anti-brane of $Q_a$.

\section{BRANES ON THE ELLIPTIC CURVE}

\subsection{Setup}

In this section we will
discuss the LG description of $B$-type of $D$ branes on the elliptic curve.
It is the simplest model with open string moduli, but nevertheless exhibits sufficient complexity for making its detailed study quite instructive.

The moduli space of the supersymmetric $\sigma$-model with the elliptic curve as its target space is locally a product of the K\"ahler and the complex structure moduli space of the two-dimensional torus. A Landau-Ginzburg formulation is possible at points in the K\"ahler moduli space where the $\sigma$-model possesses enhanced discrete symmetries. For concreteness we choose the enhanced $\BZ_3$-symmetry point in the K\"ahler moduli space, where the associated Landau-Ginzburg superpotential becomes 
\begin{equation} \label{eq:Wcubic}
  W \, = \, x_1^3+x_2^3+x_3^3-3\,a\,x_1x_2x_3 \ .
\end{equation}
The variable $a$ parametrizes algebraically the complex structure moduli space of the elliptic curve and it is related the usual flat complex structure modulus $\tau$ of the two-dimensional torus by the modular invariant $j$-function:
\begin{equation} \label{eq:Defa}
   j(\tau)\,=\,\frac{3 a (a^3+8)}{a^3-1} \ .
\end{equation}

The matrix factorizations of the cubic curve~\eqref{eq:Wcubic} for $a=0$ 
have been classified in ref.~\cite{Laza:2002}. To be specific, we present the two simplest matrix factorizations explicitly, however including $a$-dependence.
Our first example is the factorization $Q_L$, consisting of the following pair of
$3\times 3$-matrices \cite{Hori:2004ja,Brunner:2004mt}:
\begin{equation} \label{eq:MFlong}
\begin{aligned}    
    \cJ_L\, & =\, \begin{pmatrix} 
                             \alpha_1\,x_1 & \alpha_3\,x_2 & \alpha_2\,x_3 \\ 
                             \alpha_2\,x_2 & \alpha_1\,x_3 & \alpha_3\,x_1 \\ 
                             \alpha_3\,x_3 & \alpha_2\,x_1 & \alpha_1\,x_2
                        \end{pmatrix} \ , \\
    \cE_L\, & =\, \frac{1}{\alpha_1\alpha_2\alpha_3}
                        \begin{pmatrix} 
                             G_{11} & G_{22} & G_{33} \\
                             G_{23} & G_{31} & G_{12} \\
                             G_{32} & G_{13} & G_{21}
                        \end{pmatrix} \ ,
\end{aligned}
\end{equation}
with the quadratic polynomials
\begin{equation}
     G_{jk}\,=\,\alpha_{\left[k+1\right]_3} \alpha_{\left[k+2\right]_3}\,x_j^2
                  \,-\, \alpha_k^2\,x_{\left[j+1\right]_3} x_{\left[j+2\right]_3} \ .
\end{equation}
Here $[\,\cdot\,]_3$ indicates that the indices should be taken modulo three. The matrix pair, $(\cJ_L, \cE_L)$, gives rise to a valid factorization of the cubic Landau-Ginzburg superpotential $W$, as long as the complex parameters $\alpha_\ell$ obey the cubic constraint
\begin{equation} \label{eq:AlphaCubic}
     0 \, \equiv \, \alpha_1^3+\alpha_2^3+\alpha_3^3 - 3\,a\,\alpha_1\alpha_2\alpha_3 \ .
\end{equation}
Since the parameters $\alpha_\ell$ appear only in the matrix factorization but do not enter in the bulk Landau-Ginzburg superpotential~\eqref{eq:Wcubic} they parametrize the open-string moduli space of the brane described by the matrices~\eqref{eq:MFlong}. A closer look reveals that we generate a gauge-equivalent matrix factorization by homogeneously rescaling the open-string parameters $\alpha_\ell$. Therefore we can view the parameters $\alpha_\ell$ as projective coordinates of the projective space $\BC\BP^2$, which in addition are constrained to the hypersurface~\eqref{eq:AlphaCubic} \cite{Brunner:2004mt,Govindarajan:2005im}. Thus for the matrix factorization $Q_L$, we find that the open-string moduli space is just the hypersurface~\eqref{eq:AlphaCubic}, which in turn describes a two-dimensional torus.\footnote{For the matrix factorization $Q_L$ a detailed analysis reveals that due to gauge equivalences one needs to consider further identifications between the parameters $\alpha_\ell$. However, the resulting open-string moduli space is still a two-dimensional torus with the same complex structure \cite{Govindarajan:2005im}.}
 
The second example is the matrix factorization, $Q_S$, which arises from the $2\times 2$-matrix pair \cite{Brunner:2004mt}
\begin{equation} \label{eq:MFshort}
    \cJ_S\, = \,\begin{pmatrix}  
                             \phantom{-}L_1 & F_2 \\ 
                             -L_2 & F_1
                        \end{pmatrix} \ ,  \quad
    \cE_S\, = \,\begin{pmatrix} 
                             F_1 & -F_2 \\
                             L_2 &  \phantom{-}L_1
                         \end{pmatrix} \ .
\end{equation}
The linear polynomials, $L_1$ and $L_2$, are given by
\begin{equation} \label{eq:LinPoly}
     L_1 \, = \, \alpha_3 x_1 - \alpha_1 x_3  \ , \quad
     L_2 \, = \, \alpha_3 x_2 - \alpha_2 x_3 \ .
\end{equation}
If we impose again the cubic constraint~\eqref{eq:AlphaCubic} on the open-string parameters $\alpha_\ell$, then the cubic Landau-Ginzburg superpotential $W$ vanishes at the intersection locus of the two linear polynomials~\eqref{eq:LinPoly}. Due to the Nullstellensatz we can then always find two quadratic polynomials, $F_1$ and $F_2$, such that 
\begin{equation} \label{eq:Sfactors}
    W \, \equiv \, L_1F_1+L_2F_2 \ .
\end{equation} 
Then the factorization~\eqref{eq:MFshort} becomes well-defined. For the matrix factorization $Q_S$ we thus find again a toroidal open-string moduli space, parametrized by the projective coordinates $\alpha_\ell$ subject to the constraint~\eqref{eq:AlphaCubic}.

So far we have ignored an important point. In order to really describe the supersymmetric $\sigma$-model of the elliptic curve at the enhanced symmetry point in the K\"ahler moduli space, we need to consider a $\BZ_3$ orbifold of the Landau-Ginzburg theory with cubic potential. The appropriate $\BZ_3$-orbifold action on the Landau-Ginzburg fields, $x_\ell$, is generated by
\begin{equation} \label{eq:Z3action}
     x_\ell \rightarrow \omega\,x_\ell \ , 
     \quad \omega\equiv e^\frac{2\pi i}{3} \ .
\end{equation}
The topological B-branes of the resulting Landau-Ginzburg orbifold theory are now captured in terms of $\BZ_3$-equivariant matrix factorizations of the Landau-Ginzburg superpotential~\eqref{eq:Wcubic}; that is to say in order to incorporate the $\BZ_3$-orbifold action in the boundary sector, we need to enhance the matrix factorizations of the cubic superpotential $W$ to $\BZ_3$-equivariant matrix factorizations \cite{Ashok:2004zb,Walcher:2004tx,Govindarajan:2005im}. This is achieved by supplementing a given factorization, defined by $Q$, with a $\BZ_3$ representation, $R(k)$, $k\in\BZ_3$, such that the following equivariance condition is fulfilled:
\begin{equation} \label{eq:MFequi}
    Q(x) \, \equiv \, R(k)\,Q(\omega^kx)\,R(-k) \  .
\end{equation}
Note that given an equivariant matrix factorization, $(Q,R(k))$, we immediately deduce a whole $\BZ_3$-equivariant orbit of matrix factorizations distinguished by the three representations
\begin{equation}
    R^a(k)\,=\,\omega^{a k} R(k) \ , \quad a=1,2,3 \ .
\end{equation}
Hence we associate to a given (indecomposable) matrix factorization of the superpotential, $W$, a whole orbit of equivariant matrix factorizations, which differ by the $\BZ_3$-valued label, $a$, or more precisely by the $\BZ_3$-representations, $R^a(k)$.

For the two matrix factorization, $Q_L$ and $Q_S$, introduced in eqs.~\eqref{eq:MFlong} and \eqref{eq:MFshort}, we find respectively the $\BZ_3$-equivariant representations
\begin{equation}
    R^{L_a}(k) \, = \, \omega^{ak} \,
      \diag\left(\omega^{2k}\unity_{3\times 3} ,\,\unity_{3\times 3} \right) \ ,
\end{equation}
and
\begin{equation}
    R^{S_a}(k) \, = \, \omega^{ak} \,
      \diag\left(\unity_{2\times 2},\,\omega^k,\,\omega^{2k}\right)  \ .
\end{equation}

In the following we denote the brane configurations associated to these equivariant matrix factorizations by the `long' branes, $L_a$, and the `short' branes, $S_a$. This nomenclature originates from the mirror description, where these branes wrap long and short diagonals of the A-model mirror torus \cite{Brunner:2004mt}.

\subsection{Bundle geometry}

At the large radius point of the K\"ahler moduli space, quantum corrections are suppressed and therefore classical geometry yields a good description. Thus in order to get a geometric intuitive picture of topological B-branes in the Landau-Ginzburg phase, we need to translate the notion of matrix factorizations into geometric notions at the large radius point (for a detailed upcoming study, see \cite{HeHo}).
In the large radius regime topological B-branes are realized as complex vector bundles. For our purposes here the bundle picture of branes suffices, however, we should keep in mind that a more accurate description of topological B-branes at the large radius point is achieved in terms of coherent sheaves or even more precisely in terms of objects in the bounded derived category of coherent sheaves \cite{Douglas:2000gi,Lazaroiu:2001jm,Diaconescu,Aspinwall:2001pu,Distler:2002ym}.

Our goal is now to deduce directly from the matrix factorization the associated large radius bundle data of the topological B-brane. For the elliptic curve this procedure is straightforward, since the (indecomposable) complex vector bundles on the elliptic curve have been classified a long time ago \cite{Atiyah:1957}. Namely such a bundle is completely characterized by its rank $r$, its first Chern number $c_1$, as well as by a point $u$ on the elliptic curve.\footnote{Rather, on the Jacobian which happens to be isomorphic to the curve itself.} We denote such an indecomposable vector bundle by $\cV(r,c_1,u)$. The second simplification arises from the fact that at least one brane in each equivariant orbit of an indecomposable matrix factorization corresponds to a complex vector bundle in the large radius regime \cite{Laza:2002,Govindarajan:2005im}. We call this vector bundle the distinguished bundle. The other branes in the same equivariant orbit give rise to 
vector bundles of lower rank than the distinguished bundle, and/or to more general objects such as coherent sheaves. This will become somewhat more clear in a moment.

Thus our goal is to calculate for each matrix factorization of the
elliptic curve the distinguished bundle parameters $(r,c_1,u)$.  In
our two examples of factorizations given by $Q_L$ and $Q_S$, we
have already encountered an open string modulus, which was encoded
in the parameters $\alpha_\ell$ via eq.~(\ref{eq:AlphaCubic}). One
can ``uniformize'' these functions in terms of theta functions
(roughly by rewriting $\alpha\sim\theta(u|\tau)$, see
\cite{Brunner:2004mt,Govindarajan:2005im,Knapp:2007kq} for details),
and this explicitly introduces an open string modulus $u$ which is
a flat coordinate labeling a point of the (Jacobian of the) curve.
Physically this point corresponds to the location of the $D0$-brane
component of the given brane configuration.

It thus remains to determine the rank $r$ and the first Chern
number $c_1$ of the complex vector bundle. Physically these two
integers represent the large radius RR~charges of the described
brane: $(r,c_1)=(N_{D2},N_{D0})$, up to some ambiguity to be discussed
momentarily.
 
The key ingredient to determine the integers $(r,c_1)$ is to note
that the image of the matrix $\cJ$, restricted to the elliptic
curve $W\equiv 0$, encodes the relevant data of the distinguished
bundle \cite{Govindarajan:2005im}. Furthermore, due to the factorization
condition $W = \cJ\cE = \cE\cJ$, one can check that on the elliptic
curve $W\equiv 0$, the image of the matrix $\cJ$ coincides with
the kernel of the matrix $\cE$. This last property allows us to
immediately deduce the rank $r$, of the vector bundle by simply
computing the determinant of the matrix $\cE$:
\begin{equation}
    \det\cE\,\sim\,W^r \ . 
\end{equation}
The first Chern number of the vector bundle is just the first Chern
number of the induced determinant line bundle, which is obtained
by counting the number of zeros (with their multiplicities) of a
global holomorphic section. Here we construct first a global
holomorphic section of the image of the matrix $\cJ$, and then
count the number of zeros of the induced global section of the
determinant line bundle. These steps may sound a little complicated,
but they essentially just generalize the simple picture
of brane charges we gave in Section~2.  In practice this analysis
is straightforward, as will become clear by the following examples.

We consider first the distinguished bundle of the matrix factorization, $Q_L$, of the `long' branes. Using eq.~\eqref{eq:AlphaCubic} we find
\begin{equation}
    \det\cE_L\,=\,W^2 \ .
\end{equation}
Hence the distinguished bundle in question is of rank two. A global section of this rank-two vector bundle is given by the image of the two vectors $(1,0,0)^{\rm T}$ and $(0,1,0)^{\rm T}$, namely by the first two columns $v_1=(\alpha_1 x_1, \alpha_2 x_2, \alpha_3 x_3)^{\rm T}$ and $v_2=(\alpha_3 x_2, \alpha_1 x_3, \alpha_2 x_1)^{\rm T}$ of the matrix $\cJ_L$. The zeros of the induced global section of the determinant line bundle arise at points $[x_1:x_2:x_3]$ in $\BC\BP^2$, where the vectors $v_1$ and $v_2$ become linearly dependent and in addition are located on the elliptic curve $W\equiv 0$. A few steps of simple algebra reveal three such distinct points
\begin{equation}
    p_s \, = \, [\omega^s \alpha_3 : \alpha_1 : \omega^{-s} \alpha_2] \ ,
    \quad s=0,1,2 \ .
\end{equation}
Hence we deduce that the first Chern number of the bundle is three, and the matrix factorization $Q_L$ describes the distinguished bundle, $\cV(2,3,\zeta)$. It is easy to check that two different columns of the matrix $\cJ_L$ lead to the same result.

Our second example is the matrix factorization~\eqref{eq:MFshort} of the `short' branes. Again with eq.~\eqref{eq:AlphaCubic} we compute $\det\cE_S=W$, and thus we find that the factorization describes a distinguished rank-one vector bundle, {\it i.e.}, a line bundle. A global section of this line bundle is given by the image of the vector $(1,0)^{\rm T}$, or in other words by the first column, $v_1=(L_1,-L_2)^{\rm T}$, of the matrix $\cJ_S$. Here the determinant line bundle is obviously just the line bundle itself and hence we need to determine the number of points on the elliptic curve, where the vector $v_1$ vanishes. These points are given by the common zeros of the linear polynomials, $L_1$ and $L_2$, of eq.~\eqref{eq:LinPoly}. One immediately finds a single zero at
\begin{equation} \label{eq:zerolinears}
    p \, = \, [\alpha_1 : \alpha_2 : \alpha_3 ] \ .
\end{equation}
Thus we associate to the matrix factorization $Q_S$ the distinguished bundle $\cV(1,1,u)$.

Now there are a few comments in order. First of all, in the second example the attentive reader may have noticed that one obtains a different result if one constructs the global section of the image of the matrix $\cJ_S$ from the vector $(0,1)^{\rm T}$. For this choice one obtains the vector $v_1=(F_1, F_2)^{\rm T}$, and thus needs to determine the common zeros of the quadratic polynomials $F_1$ and $F_2$. Since a quadratic polynomial generically intersects the cubic polynomial $W$ in six points, the quadratics $F_1$ and $F_2$ have each six zeros on the elliptic curve. Furthermore, using the fact that the linears, $L_1$ and $L_2$, have three zeros on the elliptic curve with one common zero at the point~\eqref{eq:zerolinears}, and by carefully evaluating the relation~\eqref{eq:Sfactors} for all the zeros of the polynomials, $F_1$, $F_2$, $L_1$ and $L_2$, one finds that the polynomials $F_1$ and $F_2$ must have four common zeros. Hence from the second column of the matrix $\cJ_S$ one arrives at the distinguished bundle $\cV(1,4,u)$. 

This apparent discrepancy, however, has a simple physical resolution. In computing the $RR$~charges of the distinguished bundle, we implicitly specify a path in the K\"ahler moduli space from the Landau-Ginzburg to the large radius phase. The result depends on the choice of this path, and there is an ambiguity in the data of the distinguished bundle that is given by the monodromy about the large radius point. This monodromy amounts to tensoring the bundle data with the canonicial bundle of the ambient space $\BC\BP^2$ \cite{Laza:2002,Govindarajan:2005im,Jockers:2006sm}\footnote{Here we refer to the large radius monodromy as seen from the gauged linear $\sigma$-model point of view ({\it c.f.} ref.~\cite{Jockers:2006sm}).}\goodbreak
\begin{equation} \label{eq:LRmono}
    \cV(r,c_1,u) \xrightarrow{\ {\rm LR\ monodromy}\ } \cV(r, c_1 \pm 3 r,u) \ .
\end{equation}
Note that we can also obtain the distinguished bundle $\cV(1,4,u)$
from the first column of the matrix $\cJ_S$, by considering the
global section resulting from the image of the vector $(x_1, 0)^{\rm
T}$. Similarly we can also deduce from the matrix factorization
$Q_L$ the distinguished bundle $\cV(2,9,u)$, by computing the
global section from the image of the matrix $\cJ_L$ for the
vectors $(x_1, 0, 0)^{\rm T}$ and $(0, x_2, 0)^{\rm T}$. Thus we
can realize the large radius monodromy shifts~\eqref{eq:LRmono} by
varying the preimage sections of the matrices, $\cJ$.\footnote{By
also allowing for meromorphic preimage sections we can also obtain
negative shifts for the first Chern number.}

\begin{figure}[htb]
\hspace{\stretch{1}}
\includegraphics[width=10pc]{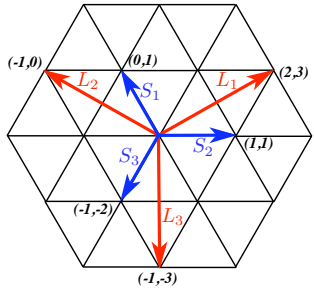}
\hspace{\stretch{1}} \\[-30pt]
\caption{In the RR-charge lattice at the $\BZ_3$-symmetric Landau-Ginzburg point in the K\"ahler moduli space, we depict the $D$-brane charges $(N_{D2},N_{D0})=(r,c_1)$ of the `long' and `short' branes $L_a$ and $S_a$, $a=1,2,3$. The distinguished bundles of the matrix factorizations $Q_L$ and $Q_S$, are associated to the branes $L_1$ and $S_2$, respectively.}
\label{fig:LSbranes}
\end{figure}
So far we have described a method to calculate on the elliptic curve the large radius RR~charges of the brane of an indecomposable matrix factorization, which is associated to the distinguished bundle. The RR~charges of the remaining branes in the equivariant orbit are then obtained by acting on them with the $\BZ_3$ monodromy associated with the ``Gepner point'' in the K\"ahler moduli space \cite{Jockers:2006sm}. The resulting large radius RR charges of all the `long' and `short' branes $L_a$ and $S_a$ have been summarized in fig.~\ref{fig:LSbranes}.

Note that the charges are not necessarily ones of nice vector bundles (\ie, of classical smooth gauge field configurations on the world-volume). For example, the $D0$ brane formally corresponds to a ``bundle'' with vanishing rank but nevertheless with non-vanishing first Chern class. Clearly, no such smooth gauge field configurations exist, and this is one of the reasons why one needs to introduce mathematical notions that are more general than vector bundles, such as sheaves, in order to properly describe such objects.

In principle the described method of evaluating the bundle data also generalizes to higher dimensional target space geometries, such as the K3 surface or Calabi-Yau threefolds. There one has to determine also higher Chern numbers by counting zeros of appropriate bundles \cite{Warner:2006private}. 
However, a closer look at a generic matrix factorization in these higher dimensional geometries reveals that it is necessary to also take into account the equivariant representations to extract the distinguished bundle data. Guided by the analysis of ref.~\cite{HeHo}, where equivariant matrix factorizations are analytically continued  to the large radius point in a systematic manner, it becomes clear that one needs to truncate the matrices $\cJ$ and $\cE$ appropriately in order to determine the correct Chern numbers. 

\subsection{Deformations}

\subsubsection{Tachyon condensation}

The matrix factorizations of the elliptic curve serve also as a good example to study topological tachyon condensation. Since the formation of bound states preserves the brane charges of the constituents, we obtain already selection rules for possible condensation processes by simply looking at the involved RR~charges. For instance from Fig.~\ref{fig:LSbranes} we infer that charge conservation allows the formation of the `long' brane $L_1$ from the `short' brane $S_2$ and the `short' anti-brane $\bar S_3$. We will demonstrate now that such a condensation process is indeed possible.

Beforehand we point out that because a whole $\BZ_3$ equivariant orbit is associated with a given matrix factorization, we need to somehow specify which precise composite out of which precise members of the two orbits under consideration we actually talk about. This extra input is provided by the choice of boundary changing operator $\Psi$ that we switch on to condense. Each channel for composite formation is characterized by a particular cohomology element, the charge of which is determined by the relative angle of the intersecting branes (in the mirror $A$-models). In other words, given two matrix factorizations, there will be a collection of boundary changing cohomology elements with different charges, and a specific choice will correspond to a tachyon consideration between specific members of the two $\BZ_3$ orbits.

Thus, if we want to form a composite out of the constituents $S_2$ and $\bar S_3$, we first need to find the relevant fermionic boundary changing operator that triggers this particular condensation process. Evaluating the BRST cohomology~\eqref{Qaction} of the matrix factorizations $Q_S$ and $Q_{\bar S}$, one finds a fermionic operator, $\Psi_{S_2\bar S_3}$, which has the following form \cite{Govindarajan:2005im}\footnote{In order to determine the equivariant labels of the boundary changing operator it is necessary to introduce an equivariant grading for the BRST cohomology elements (for details, {\it c.f.} refs.~\cite{Ashok:2004zb,Walcher:2004tx,Govindarajan:2005im}).}
\begin{equation}
    \Psi_{S_2\bar S_3}\,=\,
    \begin{pmatrix}
        0 & \psi_0(u_{S_2},u_{\bar S_3}) \\ \psi_1(u_{S_2},u_{\bar S_3}) & 0
    \end{pmatrix} \ .
\end{equation}
It depends on the open-string moduli, $u_{S_2}$ and $u_{\bar S_3}$, of both constituents, $S_2$ and $\bar S_3$. The matrices, $\psi_0$ and $\psi_1$, schematically read:
\begin{equation} \label{eq:TorusBdryChOp}
\begin{aligned}     
     \psi_0(u_{S_2},u_{\bar S_3})\,&=\,
       \begin{pmatrix}
            l_5(x) & q(x) \\ c & l_6(x)
       \end{pmatrix} \ , \\
      \psi_1(u_{S_2},u_{\bar S_3})\,&=\,
       \begin{pmatrix}
            l_1(x) & l_2(x) \\ l_3(x) & l_4(x)
       \end{pmatrix} \ .
\end{aligned}
\end{equation}
Here $c$ is a constant, $l_i(x)$ and $q(x)$ are linear and quadratic polynomials in the variable, $x$. In addition all entries have non-polynomial dependencies on the open-string moduli, $u_{S_2}$ and $u_{\bar S_3}$. 

If we now employ the condensation formula~\eqref{eq:MFcone},
we obtain for the resulting matrix factorization, $Q_C$, the $4\times 4$ matrix pair
\begin{equation}
\begin{aligned}
    \cJ_{\rm C} \,&=\, 
    \begin{pmatrix}
         \cE_S(u_{\bar S_3}) & \psi_0(u_{S_2},u_{\bar S_3}) \\
         0 & \cJ_S(u_{S_2})
    \end{pmatrix} \ , \\
    \cE_{\rm C} \,&=\,
    \begin{pmatrix}
         \cJ_S(u_{\bar S_3}) & \psi_1(u_{S_2},u_{\bar S_3}) \\
         0 & \cE_S(u_{S_2})
    \end{pmatrix} \ .
\end{aligned}
\end{equation}
A straightforward but tedious calculation reveals that this matrix pair is gauge equivalent to \cite{Govindarajan:2005im}
\begin{equation}
\begin{aligned}
    \cJ_{\rm C} \,&=\, 
    \begin{pmatrix}
         1 & 0 \\
         0 & \cJ_L(u_{S_2}-u_{\bar S_3})
    \end{pmatrix} \ , \\
    \cE_{\rm C} \,&=\,
    \begin{pmatrix}
         W & 0 \\
         0 & \cE_L(u_{S_2}-u_{\bar S_3})
    \end{pmatrix} \ .
\end{aligned}
\end{equation}
Hence by eliminating a trivial brane-anti-brane pair, we find that
the resulting matrix factorization is indeed equivalent to the
matrix factorization $Q_L$ of the `long' brane $L_1$. The appearance
of the trivial brane-anti-brane pair can be traced back to the
constant entry $c$ in the fermionic boundary changing
operator~\eqref{eq:TorusBdryChOp}, as it allows for a chain of gauge
transformations that eventually makes the trivial brane-anti-brane
pair manifest\cite{Govindarajan:2005im}. Moreover, we observe that
the condensate depends again only on a single open string modulus
given by $u_L\equiv u_{S_2}-u_{\bar S_3}$. This is in agreement
with the general property of indecomposable matrix factorizations
on the elliptic curve, which depend always just on a single open-string
modulus \cite{Atiyah:1957,Laza:2002}.

A known feature of the elliptic curve is the fact that any
(indecomposable) matrix factorization can be obtained through
successive tachyon condensation out of `short' branes
\cite{Laza:2002,Govindarajan:2005im}, so these serve as generators
of the $D$-brane charge lattice. One certainly expects similar
features to hold also for higher dimensional Calabi-Yau's.

\subsubsection{Effective superpotential}

By solving the cohomology problem for any given factorization, one
can obtain explicit expressions for the moduli-dependent open string
vertex operators. With those at hand, one may want to go on and
compute correlation functions, for example ones that give rise to
a superpotential on an intersecting brane configuration.  Actually,
computing correlators with more than three boundary changing operator
insertions is hard, because they necessarily involve integrated
insertions and there is no simple method to evaluate them except
in favorable circumstances. So let's focus on three-point functions
for three intersecting `long diagonal' branes on the cubic elliptic
curve for the time being \cite{Brunner:2004mt} (see also
\cite{Knapp:2007kq}). Note that the `long diagonal' branes intersect
three times on the curve (in the $A$ model picture), so there are
three boundary changing operators $\Psi_{ab}^i$, $i=1,2,3$, in the
open string spectrum between any pair of those branes, one for each
intersection point. The correlators thus have the following form:
\beql{threepoint}
C_{ijk}\ =\ \langle\,\Psi_{12}^i\Psi_{23}^j\Psi_{31}^k\,\rangle\ ,
\eeq
and the issue is to compute their dependence on the closed and open
string moduli.  Matrix representatives for the $\Psi's$ were
explicitly given in refs.~\cite{Brunner:2004mt,Govindarajan:2005im},
and the simplest method to evaluate the correlators is to just plug
these matrices into the Kapustin-Li \cite{Kapustin:2002bi,Herbst:2004ax}
supertrace residue formula:
\beql{residue}
\langle\Psi_{12}^i\Psi_{23}^j\Psi_{31}^k\rangle =
\!\int \!\frac{d^3x}{\prod dW(x)}{\rm Str}\!
\left[ dQ^{\wedge 3}\Psi_{12}^i\Psi_{23}^j\Psi_{31}^k\right]\!,
\eeq
and evaluate it in a straightforward manner. 
One then obtains explicit expressions for
the correlators (\ref{threepoint}), which depend on the parameters $a$ and
$\alpha_\ell$ via the various matrices.

However the tricky part is to determine the correct  normalization
of the vertex operators: it is in general moduli dependent, and
thus the computation done so far is seriously incomplete. Determining
the normalization factors is actually the main part of the work.  As it turns out
\cite{WL}, there exist certain differential equations that determine
these normalization factors, but discussing those here would lead too
far outside the scope of this lecture.  At any rate, for the elliptic
curve there are other, indirect arguments to infer the correct
normalization \cite{Brunner:2004mt}.

After we have determined the correct normalization factors in one way or
other, it is then at this point more interesting to switch to the
topological $A$-model picture. This is done by employing the mirror
map, which maps the $B$-model complex structure parameters $a$,
$\alpha_\ell$ to flat coordinates $\tau$, $u$. By mirror symmetry
the latter coincide with the K\"ahler variables of the $A$-model.
As already mentioned, this map is given by (\ref{eq:Defa}) for the
closed string modulus, and by certain theta functions for the open
string moduli:
\beql{alphadef}
\alpha_\ell(u,\tau)\ =\ \omega^{(\ell -1)}  \, 
\Theta \bigg[{ {{1 \over 3} (1-\ell) -{1 \over 2} \atop -{1 \over 2}} }\, \bigg |\,3\,u,3
\,\tau \bigg]\ ,\ \ \
\eeq
where $\ell=1,2,3$, $\omega=e^{2\pi i/3}$, $q=e^{2\pi i\tau}$ and
\beql{thetadef}
\Theta \bigg[{ {c_1 \atop c_2}} \bigg  |\,u \,, \tau\bigg]\ =\
\sum_{m \in \BZ}  q^{(m+c_1)^2/2} e^{2\pi i(u +c_2)(m+c_1)}\,.
\eeq
One can easily check that $a(\tau)$ and the $\alpha_\ell(u,\tau)$ as given here satisfy the cubic
constraint (\ref{eq:AlphaCubic}).

\begin{figure}[htb]
\vspace{2pt}
\includegraphics[width=20pc]{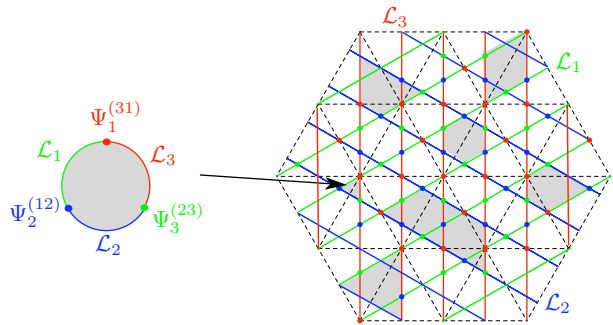}
\caption{
In the $A$-model mirror picture, the three-point function gets contributions
from triangle shaped instantons that are bounded by three, pairwise intersecting
$D1$ branes. As shown, the instantons correspond to holomorphic
maps from the world-sheet disk into the target space $D$-brane
geometry
(we drew it here on the covering space of the torus).  Similarly,
higher point correlators get contributions from polygonal instantons,
some examples of which are shown as well. For the elliptic curve they can be evaluated directly in the $A$-model \cite{Polishchuk:2000kx,Herbst:2006nn,Knapp:2007kq,Kajiura:2007zz}. How to explicitly
compute them for general Calabi-Yau manifolds is an open problem.}
\label{fig:instantons}
\end{figure}

After taking all normalization factors properly into account, the Yukawa 
couplings (\ref{threepoint}) then eventually come out as follows:
\beqal{finalcorr}
C_{111}(\tau,u_i) &=& \alpha_1(u_1+u_2+u_3,\tau)\non\\
C_{123}(\tau,u_i) &=& \alpha_2(u_1+u_2+u_3,\tau)\\
C_{132}(\tau,u_i) &=& \alpha_3(u_1+u_2+u_3,\tau)\non
\eeqal
(plus obvious cyclic transforms of the indices).  This reproduces
results first obtained in \cite{Cremades:2003qj,Cremades:2004wa}.
Besides potentially interesting for phenomenological model building,
the $A$-model expressions (\ref{finalcorr}) are interesting also
for other reasons: Under mirror symmetry, the $B$ type $D0$,$D2$-branes
map into $A$ type $D1$ branes wrapping special lagrangian 1-cycles, and the
interpretation of the $q$ series expansion is in terms of disk
instantons that span between the three intersecting $D1$ branes in
question -- see Fig.~\ref{fig:instantons}.

Having computed these expansions directly from the topological
$B$-model means that they represent a test of {\it homological
mirror symmetry} \cite{kontsevich,Polishchuk:1998db} (see \cite{Knapp:2007kq}
for similar computations for the quartic curve).  Particularly
interesting would be a generalization to higher dimensional, compact
Calabi-Yau manifolds, as the issue of homological mirror symmetry
becomes for those much more non-trivial. We intend to report progress
on this elsewhere.

Suffice it for the time being to approach the problem of computing
correlators also from a different angle, namely via the deformation
theoretical method outlines in Sect.~3.1.  This method is suitable
for computing higher than three-point functions, but we will restrict
here ourselves, for illustrative purposes, to briefly recover the
previous result for the three point function in that language.

We consider the following BRST operator, which takes the brane
geometry into account: 
\beq 
Q=
\begin{pmatrix}
Q_1&T_{12}^i\Psi_{12}^i&\\
&Q_2&T_{23}^i\Psi_{23}^i\\
T_{31}^i\Psi_{31}^i&&Q_3\\
\end{pmatrix}\ .
\eeq 
We have seen in the previous section
that when we switch on just a single tachyon VEV, factorization is
unspoiled and thus, this deformation in unobstructed. However if
we now switch on two tachyons, there is an obstruction at
second order, \ie:
\beql{obstr}
Q^2-W\ =\ T_{12}^iT_{23}^j \,\lambda_2(\Psi_{12}^i,\Psi_{23}^j)\ .
\eeq
This is because in the OPE
\beql{OPE}
\lambda_2 (\Psi_{12}^i,\Psi_{23}^j)\ \equiv\ \Psi_{12}^i\cdot\Psi_{23}^j\ =\ 
C_{ij}^k\,\Phi_{13;k} + \big\{Q_{13},*\big\}
\eeq
there appears non-trivial bosonic cohomology elements, $\Phi_{13;k}\in
{\rm Ext}^2(P_1,P_3)$, on the  RHS. This means that the iterative
procedure stops already at second order. The boundary ring
structure constants, $C_{ij}^k$, essentially coincide with the
three-point correlators given above (again, correct normalization
of all operators is essential here). To show this, it is crucial
to make use of certain theta-function identities. These are of the
generic form \cite{Polishchuk:1998db}
\beq
\Theta(u_1,\tau)\cdot\Theta(u_2,\tau)\sim \sum \Theta(u_1-u_2,\tau)\cdot\Theta(u_1+u_2,\tau) ,
\eeq
which mirrors the structure of the OPE (\ref{OPE}) (such identities actually encode the products in the relevant Fukaya category).

As claimed in Section 3.1, the obstruction given by the RHS of (\ref{obstr})
determines a derivative of the effective
superpotential. From (\ref{Weff}) the latter is given (to third order) by
\beqal{wefftau}
\weff(\tau,u)&\sim&  T_{12}^iT_{23}^j \langle\,T_{31}^\ell\Psi_{31}^\ell,\,
 \Pi\lambda_2(\Psi_{12}^i,\Psi_{23}^j)\rangle
 \non\\
&=&  T_{12}^iT_{23}^jT_{31}^k C_{ijk}(\tau,u)\ ,
\eeqal
where we have made use of Serre duality, \ie,
$\langle\Psi_{ab}^\ell,\Phi_{cd;k}\rangle=\delta_{k\ell}\delta_{ad}\delta_{bc}$. 
Its derivative with respect to
$T_{31}$ indeed yields the obstruction term in (\ref{obstr}) (after forming a scalar expression).


\bibliographystyle{utphys}
\bibliography{bibliography}

\end{document}